# A Sinc Wavelet Describes the Receptive Fields of Neurons in the Motion Cortex

Stephen G. Odaibo

M.S.(Math), M.S.(Comp. Sci.), M.D.

Quantum Lucid Research Laboratories

stephen.odaibo@qlucid.com

# ABSTRACT

Visual perception results from a systematic transformation of the information flowing through the visual system. In the neuronal hierarchy, the response properties of single neurons are determined by neurons located one level below, and in turn, determine the responses of neurons located one level above. Therefore in modeling receptive fields, it is essential to ensure that the response properties of neurons in a given level can be generated by combining the response models of neurons in its input levels. However, existing response models of neurons in the motion cortex do not inherently yield the temporal frequency filtering gradient (TFFG) property that is known to emerge along the primary visual cortex (V1) to middle temporal (MT) motion processing stream. TFFG is the change from predominantly lowpass to predominantly bandpass temporal frequency filtering character along the V1 to MT pathway (Foster et al 1985; DeAngelis et al 1993; Hawken et al 1996). We devised a new model, the sinc wavelet model (Odaibo, 2014), which logically and efficiently generates the TFFG. The model replaces the Gabor function's sine wave carrier with a sinc (sin(x)/x) function, and has the same or fewer number of parameters as existing models. Because of its logical consistency with the emergent network property of TFFG, we conclude that the sinc wavelet is a better model for the receptive fields of motion cortex neurons. This model will provide new physiological insights into how the brain represents visual information.

KEYWORDS: Visual Motion, Receptive Field, Gabor wavelet, Sinc wavelet

# INTRODUCTION

As one traces the neural pathway connecting the input layers of V1 to area MT, neurons become increasingly specialized for motion detection. In ascending order, attributes such as orientation selectivity, direction selectivity, speed tuning, increasing preferred speeds, component spatial frequency selectivity, and pattern (plaid) spatial frequency selectivity are sequentially acquired. Furthermore, the temporal frequency filtering properties of the neuronal population changes in a particular way along this pathway. Specifically, the proportion of bandpass temporal frequency filtering neurons to lowpass temporal frequency filtering neurons increases. We term this the Temporal Frequency Filtering Gradience (TFFG) property. Existing receptive field models do not represent this fundamental emergent property. Our specific aim is to introduce and describe a model for the receptive fields of V1 to MT neurons which innately and efficiently represents the aforementioned properties.

Kuffler's early studies of retinal ganglion cell response properties led to similar studies of simple cortical cells by Hubel and Wiesel. Together, their work generated great interest in neuronal receptive fields (Daugman, 1985; Hubel & Wiesel, 1959; Hubel, 1959; Hubel, 1957; Kuffler, 1953; Marčelja, 1980). The functional forms of these receptive fields are at once beautifully simple yet enormously complex. Hence mathematical models have been used in step with electrophysiological studies to advance our understanding of their properties. Initially, focus was predominantly on their spatial structure (Gabor, 1946; Marčelja, 1980). Now, however, their temporal structure is increasingly studied in tandem. In particular, for most neurons in the V1 to MT processing stream, it is now appreciated that spatial and temporal features cannot be studied separately. They are spatiotemporally inseparable entities. In particular, motion is encoded by orientation in the spectral domain, and spatiotemporally-oriented filters are therefore motion detectors (Heeger, 1987; Watson & Ahumada Jr, 1983; Watson & Ahumada Jr, 1985). Therefore at first glance it may seem an easy matter to mathematically model motion detecting neurons. The challenge, however, is to develop physiologically sound receptive field models which reflect the hierarchical structure of the motion processing stream. Various models do exist which are spatiotemporally-oriented filters, and are therefore motion detectors from a mathematical standpoint (Adelson & Bergen, 1985; Qian, 1994; Qian & Ersen, 1997; Qian, Andersen, & Adelson, 1994; Qian & Freeman, 2009). However, they fail to represent one of the most salient characterizing attributes of the motion processing stream: the lowpass to bandpass distribution of temporal frequency filtering properties along the V1 to MT specialization hierarchy. Next we discuss one of the most fundamental of such emergent network properties: the particular distribution of temporal frequency filtering types along the stream (DeAngelis, Ohzawa, & Freeman, 1993; Foster, Gaska, Nagler, & Pollen, 1985; Hawken, Shapley, & Grosof, 1996).

Hawken et al found that direction-selective cells were mostly bandpass temporal frequency filters, while cells which were not direction-selective were equally distributed into bandpass and lowpass temporal frequency filtering types (Hawken et al., 1996). Foster et al found a similar phenomenon in macaque V1 and V2 neurons. V1 neurons were more likely to be lowpass temporal frequency filters, while their more specialized downstream heirs, V2 neurons, were more likely to be bandpass temporal frequency filters (Foster et al., 1985). Less specialized neurons located anatomically upstream (lower down in the hierarchy) are more likely to have lowpass temporal frequency filter characteristics, while more specialized neurons located anatomically downstream (higher up in the hierarchy) are more likely to display bandpass temporal frequency filter characteristics. For example, LGN cells are at best only weakly tuned to direction and orientation (Ferster & Miller, 2000; Xu, Ichida, Shostak, Bonds, & Casagrande, 2002) and are hence equally distributed into lowpass and bandpass categories. Layer 4B V1 cells and MT cells on the other hand, are almost all direction selective (Maunsell & Van Essen, 1983) and hence are mostly bandpass temporal frequency filters. We firmly believe this classification is not arbitrary, but instead is a direct manifestation of the particular spatiotemporal structure of the V1 to MT motion processing stream. Consequently, the receptive field model must reflect this increased tendency for bandpass-ness with ascension up the hierarchy. In other words, the representation scheme must be one that is inherently more likely to deliver bandpass-ness to a more specialized cell (such as in layer 4B or MT) and lowpass-ness to a less specialized cell (such as in layer 4A, 4Cα, or 4Cβ). However, none of the existing receptive field models reflect this underlying spatiotemporal structure. In contrast, as will be seen in the Methods section below, the Sinc wavelet's wave carrier is a sinc function which directly confers the TFFG property. Individual Sinc wavelet basis elements are lowpass temporal frequency filters; and bandpass temporal frequency filters can only be obtained by combinations of basis elements. A study by DeAngelis and colleagues also corroborates the above. They found that temporally monophasic V1 cells in the cat were almost always low pass temporal frequency filters, while temporally biphasic or multiphasic V1 cells were almost always bandpass temporal frequency filters (DeAngelis et al., 1993). The explanation for this finding is inherent and explicit in the Sinc wavelet basis, where bandpass temporal character necessarily results from biphasic or multiphasic combination. All monophasic elements on the other hand, are lowpass temporal frequency filters. However, we will see that according to the model, the converse is not true: i.e. not all lowpass temporal frequency filters are monophasic, and not all multiphasic combinations yield bandpass temporal frequency filters.

# METHODS

The sinc wavelet differs from the standard spatiotemporal Gabor wavelet in the wave-carrier. Where the standard Gabor's wave-carrier is a sine function, the sinc wavelet's wave-carrier is a sinc (i.e. sin(x)/x) function. We define the Sinc wavelet as follows,

$$G(t,x,y) = A\exp\left(\frac{-(x-x_0)^2}{2\sigma_x^2} - \frac{(y-y_0)^2}{2\sigma_y^2}\right) sinc(\omega_0 t - u_0 x - v_0 y + \varphi)$$

**Equation 1**

where the constant multiplier $A$ is amplitude; $\sigma_x$ and $\sigma_y$ are the gaussian variances in the $x$ and $y$ directions; $x_0$ and $y_0$ are the respective $x$ and $y$ coordinates of the gaussian center; $\omega_0$, $u_0$, and $v_0$ are the frequencies of the wave carrier in the $t$, $x$, and $y$ directions respectively; $\varphi$ is the sinusoid phase; and the *sinc* function is defined as,

$$sinc(x) = \frac{\sin(x)}{x}$$

**Equation 2**

We have rotated our coordinates by an angle $\theta$ from a reference state $(x',y')$ to a state $(x,y)_\theta$ which we denote $(x,y)$ for notational simplicity. The transformation,

$$(x',y') \longrightarrow (x,y)_\theta = (x,y),$$

**Equation 3**

is given by,

$$\begin{pmatrix} x \\ y \end{pmatrix} = \begin{pmatrix} \cos(\theta) & -\sin(\theta) \\ \sin(\theta) & \cos(\theta) \end{pmatrix} \begin{pmatrix} x' \\ y' \end{pmatrix}.$$

**Equation 4**

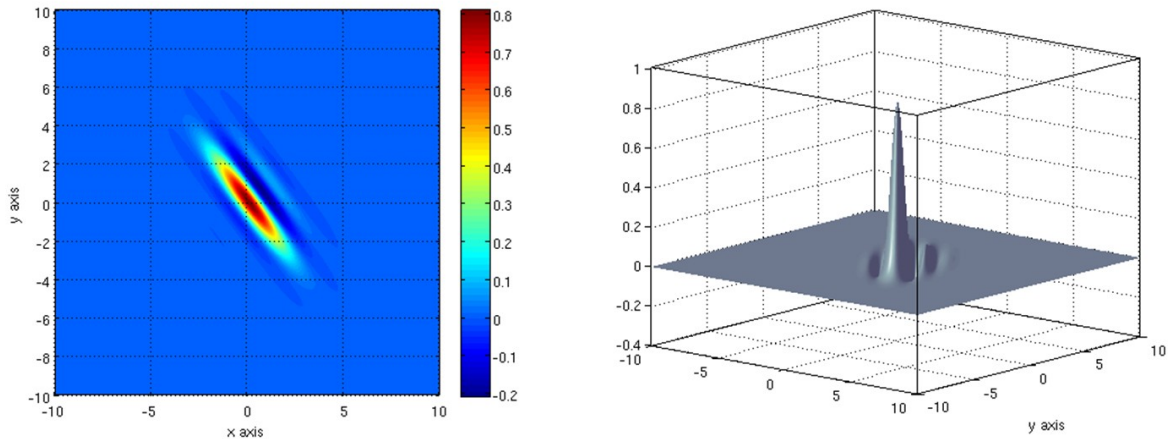

**Figure 1: An instance of the *Sinc wavelet*. Colormap on right and elevation surface plot on left.**

We proceed here with the following instance of the sinc wavelet,

$$G(t,x,y) = \exp\left(\frac{-x^2}{2\sigma_x^2} - \frac{y^2}{2\sigma_y^2}\right) \operatorname{sinc}(t - u_0 x - v_0 y)$$

**Equation 5**

where we have set $\omega_0$ equal to one. We can always do so for one reference neuron by simply defining the unit of time as $1/\omega_0$, the duration the neuron's frame cycle. As we will see, this value, $1/\omega_0$, is the shortest frame duration to which the neuron can respond. In the above equation, we have also set $A=1$, and $\theta = x_0 = y_0 = 0$.

The fourier transform is as follows,

$$H(\omega,u,v) = \frac{\sigma_x \sigma_y A}{2}\sqrt{\frac{\pi}{2}} \exp\left(\frac{-\sigma_x^2(u+u_0\omega)^2}{2} - \frac{\sigma_y^2(v+v_0\omega)^2}{2}\right) L(\omega)$$

**Equation 6**

where $L(\omega)$ is given by,

$$L(\omega) = sign(1-\omega) + sign(1+\omega)$$

**Equation 7**

and the *sign* function is defined as,

$$sign(x) = \begin{cases} 1 & if\ x>0 \\ 0 & if\ x=0 \\ -1 & if\ x<0 \end{cases}$$

**Equation 8**

The above fourier transform was obtained using Mathematica symbolic software. The fourier transform of the general case will likely be challenging to obtain analytically either by hand or symbolic software. Hence, we anticipate numerical methods may have an important role to play.

The maximum magnitude of the response function, Equation 6, is attained where the argument of the exponent is zero, i.e. where,

$$\sigma_x^2(u+u_0\omega)^2 + \sigma_y^2(v+v_0\omega)^2 = 0$$

**Equation 9**

The neuron's preferred spatial frequency, $f_0$, is dependent on the temporal frequency, $\omega$, of the stimulus, and is given by the above bivariate quadratic equation's solution,

$$f_0(\omega)=(-u_0\omega,-v_0\omega).$$

### Equation 10

The location where the response maximum is attained is a parametrized curve in 3D frequency space. It is given by,

$$R(\omega)=(\omega,-u_0\omega,-v_0\omega).$$

### Equation 11

Although the location in *(u,v)* space where the neuron's maximum response is attained depends on *ω*, the magnitude of the maximum response is itself a constant. It is given by,

$$R_{max}=|H(\omega,f_0(\omega))|=\frac{\sigma_x\sigma_y A}{2}\sqrt{\frac{\pi}{2}}.$$

### Equation 12

The half magnitude response is attained at values of *(u,v)* satisfying,

$$\exp\left(\frac{-\sigma_x^2(u+u_0\omega)^2}{2}-\frac{\sigma_y^2(v+v_0\omega)^2}{2}\right)=\frac{1}{2}.$$

### Equation 13

Taking the natural logarithm of the above equation yields,

$$\frac{-\sigma_x^2(u+u_0\omega)^2}{2\ln(2)}-\frac{\sigma_y^2(v+v_0\omega)^2}{2\ln(2)}=1.$$

### Equation 14

Defining $a:=\sigma_x^{-1}\sqrt{2\ln(2)}$ and $b:=\sigma_y^{-1}\sqrt{2\ln(2)}$, Equation 14 becomes,

$$\frac{-(u+u_0\omega)^2}{a^2} - \frac{(v+v_0\omega)^2}{b^2} = 1.$$

**Equation 15**

In the above form, one readily recognizes this as the ellipse centered at $(-u_0\omega, -v_0\omega).$ , whose principal axes radii are *a* and *b* in the *x* and *y* directions respectively. The long axis in the frequency domain is the short axis in the spatial domain and vice versa. Without loss of generality, we can assume that prior to rotation of the spatial axes by angle $\theta$, the long axis of the receptive field envelope is parallel to the *x* axis. Then the orientation of the on-off bands, i.e. planes of the spatial wave, are aligned parallel to the long axis of the envelope when the rotation angle, $\theta$, is related to the polar angle $\mu$ by the relationship,

$$\theta = \mu$$

**Equation 16**

On the other hand, the on-off bands are perpendicular to the long axis of the envelope when,

$$\mu = \theta + \frac{1}{2}$$

**Equation 17**

In the case of Equation 16, the half magnitude frequency bandwidth is readily seen to equal *2a*, the length of the short axis, while in the case of Equation 17, the half magnitude frequency bandwidth equals *2b*, the length of the short axis. The cases of skewed alignment take on values between *2a* and *2b* and are also computable from the geometry. Unlike the preferred spatial frequency, the half-magnitude frequency bandwidth is not dependent on the temporal frequency of the stimulus.

In summary, the Sinc wavelet spatiotemporal receptive field model predicts that the magnitude of a V1 to MT neuron's preferred spatial frequency is linearly dependent on the temporal frequency of the stimulus as shown in Equation 10. In the next section, we present some receptive field simulations using the Sinc wavelet. Using the above equations, numerical simulations were done in MATLAB and are shown below.

## RESULTS

Here we present simulation results which demonstrate how the sinc wavelet model innately represents the temporal frequency filtering property distribution along the V1 to MT neuronal hierarchy. The following notations are used in the figure captions: $\theta_e$ is the angle of rotation of the axes of the gaussian envelope as described in Equation 3 and Equation 4. $\theta_s$ is the angle of rotation of the axes of the wave carrier. *ctr* is the 2-component vector consisting of the *x* and *y* coordinates of the gaussian envelope center respectively. $\sigma=(\sigma_x,\sigma_y)$ is the 2-component vector consisting of the gaussian variances of the envelope in the *x* and *y* directions respectively. $\varphi$ is phase of the wave carrier. *t* is time. And $\hat{\omega}=(\omega_0,u_0,v_0)$ is the 3-component vector consisting of the temporal, *x*-spatial, and *y*-spatial frequencies of the wave carrier respectively. For succinctness, $\hat{\omega}=1$, for instance, is taken to be equivalent to, $\hat{\omega}=(1,1,1)$. The same short-hand notation applies to the other multi-component vectors.

Figure 2 shows the essential property which the Sinc wavelet inherits from the sinc function. The sinc function's fourier transform is a lowpass filter.

Higher frequency sinc functions have wider bandwidth. Bandpass filters are formed by taking the difference between sinc functions of different frequency as illustrated in Figure 2. The Sinc wavelet's fourier transform has a step function factor along the temporal frequency direction. It inherits this step functionality from the sinc function wave carrier. This allows it describe the particular distribution of lowpass to bandpass temporal frequency filter properties of V1 to MT neurons (DeAngelis et al., 1993; Foster et al., 1985; Hawken et al., 1996) in a manner innately representative of the motion-processing stream's neuronal hierarchy.

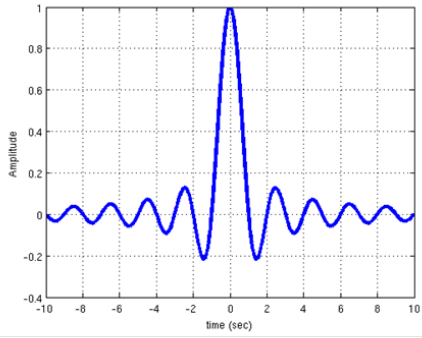
(a) Sinc, $\omega_0 = 1$

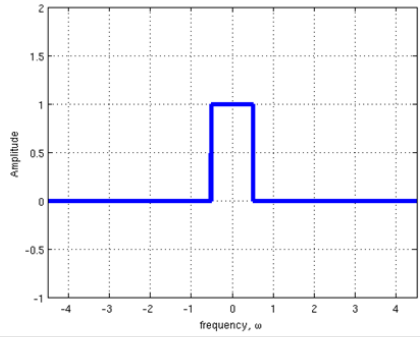
(b) "Fourier of (a)"

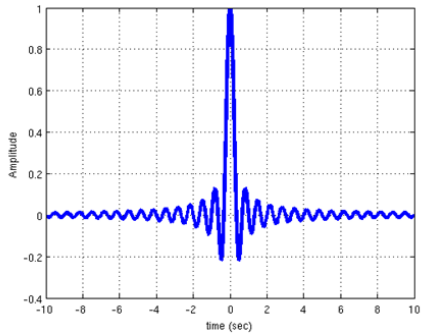
(c) Sinc, $\omega_0 = 3$

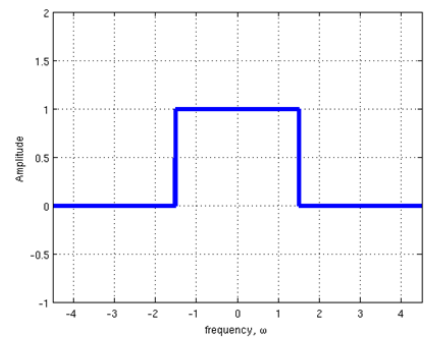
(d) "Fourier of (c)"

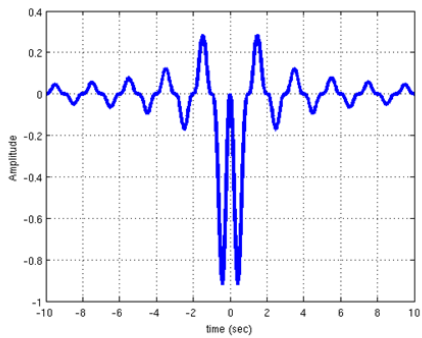
(e) "(c) minus (a)"

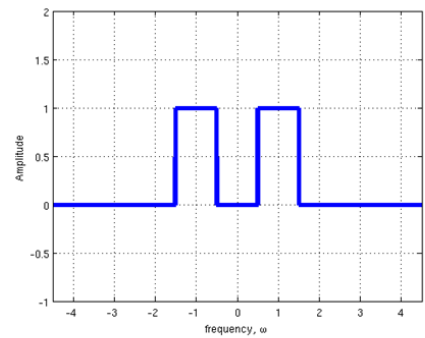
(f) "(d) minus (b)"

Figure 2: The sinc function and its fourier transform

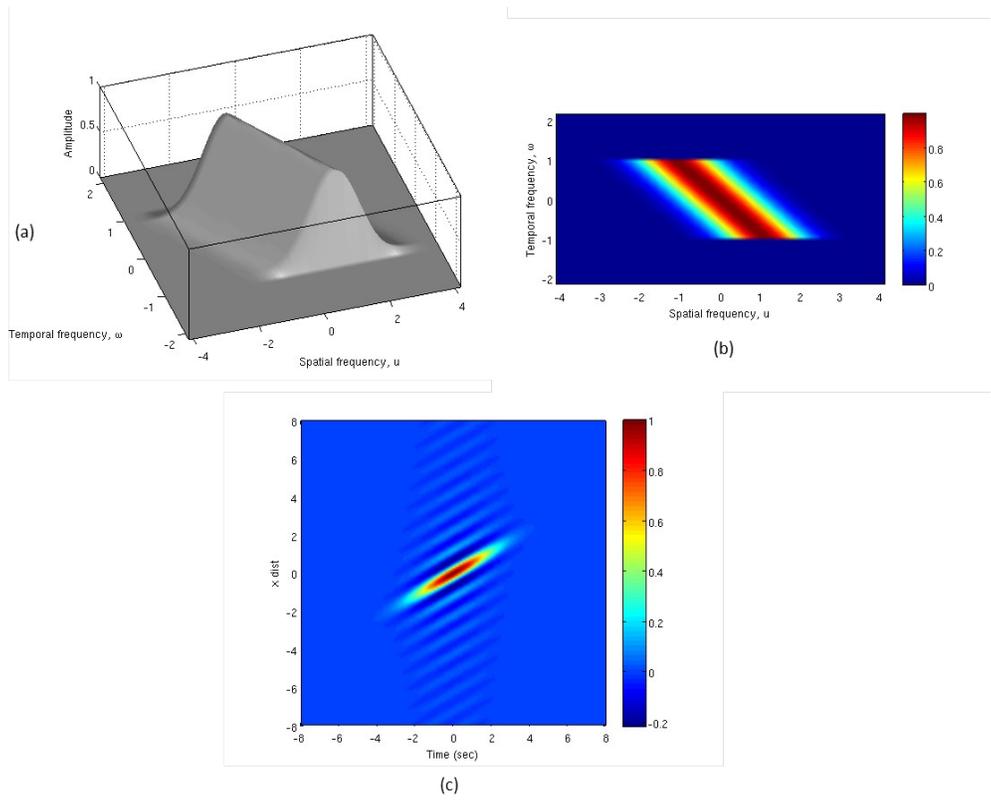

**Figure 3:**

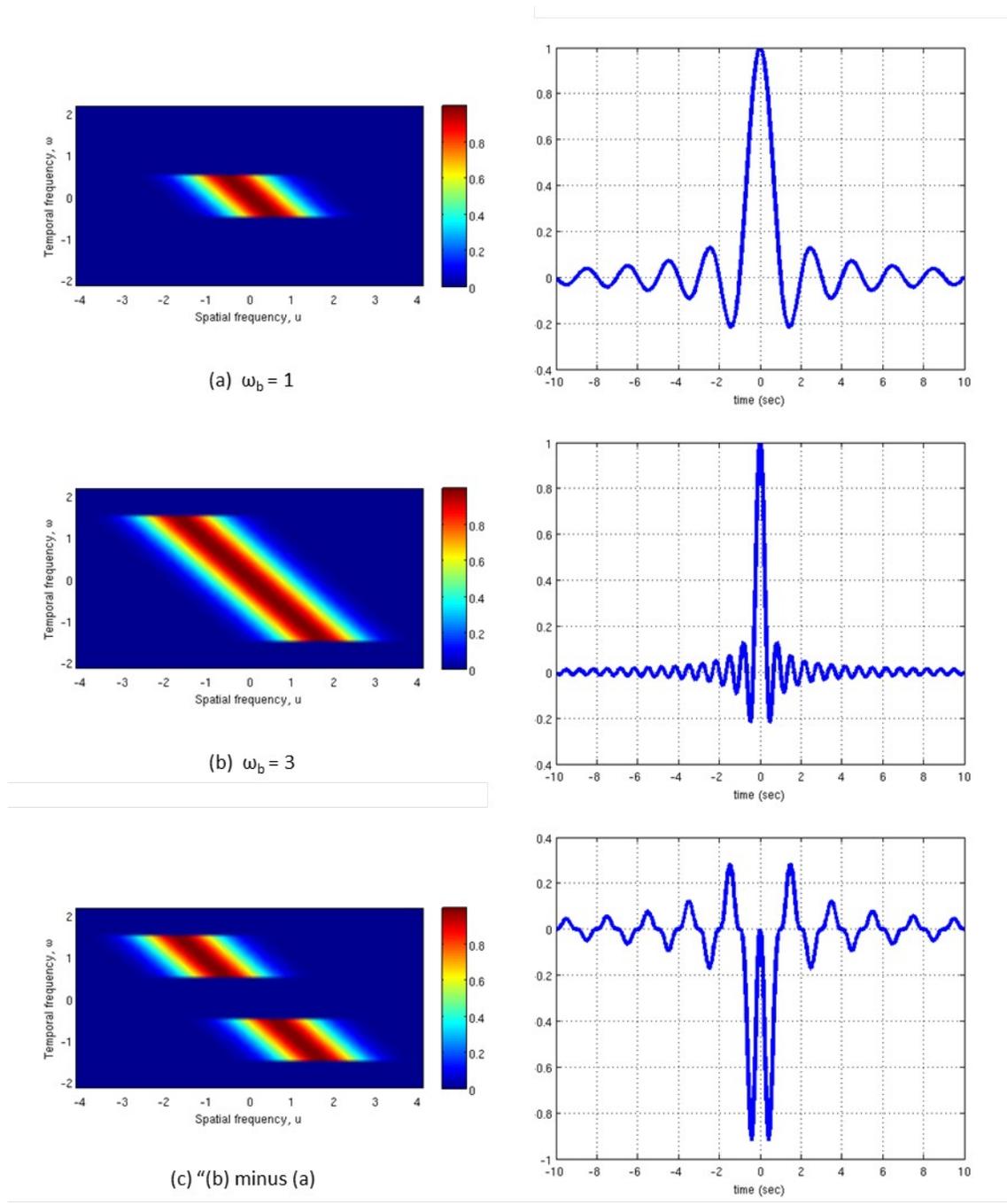

**Figure 4: Progressive complexity:** The bandpass temporal frequency filters such as (c) are necessarily superpositions of two or more sinc wavelets. The profiles on the left are the temporal response weights probed at the spatial origin. Note that the pattern is more complex in the bandpass temporal frequency filter. Parameters used: Parameters used: σ=1, u₀=1.15

For simplicity of illustration, Figure 3 shows a 2D (x,t) sinc wavelet basis element. Its fourier transform is shown in Figure 3a and Figure 3b, while its (x,t) domain representation is shown in Figure 3c. The lowpass temporal frequency nature of the fourier representation is apparent, as the support straddles the zero. Figure 4 shows the construction of sinc wavelets which are temporal frequency bandpass filters.  As illustrated, the temporal frequency bandpass sinc wavelet element results from the difference of two temporal frequency lowpass SInc wavelet basis elements. The right-hand column of Figure 4 plots sinc basis elements, while the right hand column plots the corresponding temporal response probed at the spatial origin of the receptive field. There is an increasing complexity of the temporal waveform structure as one progresses from lowpass to bandpass basis element.

Figure 4a is a plot of Equation 6, i.e. it is the fourier transform of the sinc wavelet described by Equation 5. We label this basis element "$\omega_b=1$", meaning it is the zero-centered lowpass temporal frequency filter whose temporal frequency bandwidth is one. Accordingly, Figure 4Equation 4b is a plot of the "$\omega_b=3$" sinc wavelet basis element, i.e. it is a plot of the zero-centered lowpass temporal frequency filter whose temporal frequency bandwidth is three. Figure 4c plots the bandpass temporal frequency filter basis element obtained by taking the difference of the "$\omega_b=3$" and "$\omega_b=1$" basis elements.

All pure sinc wavelet basis elements are lowpass temporal frequency filters. On the other hand, the bandpass temporal frequency filter property necessarily results from summation of sinc basis elements. Hence bandpass temporal elements are necessarily complex (not pure). However, not all lowpass temporal frequency filters are pure; and not all basis summations yield bandpass temporal frequency filters. For instance, Figure 5 shows the sum of two pure sinc basis elements which yield another lowpass temporal frequency filter. The temporal waveform of the composite element is indeed more complex than that of its two pure constituents; however, it appears less complex than the composite temporal waveform of Figure 4.

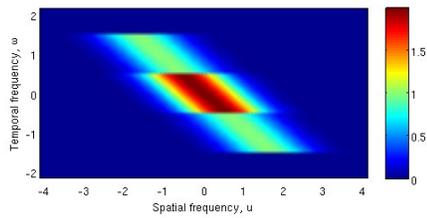
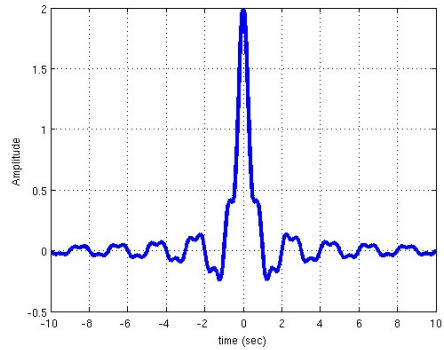
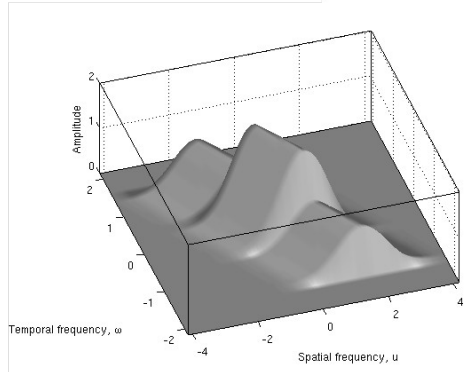

**Figure 5: A superposition of two lowpass filters which yields a hierarchically more complex neuron which is nonetheless low pass. Specifically, the above is a plot of "$\omega_b=1$" + "$\omega_b=3$". Parameters used: σ=1, $u_0$=1.15.**

## DISCUSSION

As one traverses the motion processing stream from V1 to MT, the neuronal population consists of an increasing proportion of neurons with temporal frequency bandpass filter behavior. Consequently, the proportion of temporal frequency lowpass filter neurons progressively decreases. In this paper, we have termed this the temporal frequency filtering gradient (TFFG) property. The role for this TFFG property in visual motion perception is unclear at the moment. One possibility is that bandpass filtering allows for highly specific parsing of the kinetic components of a visual scene. It is reasonable to expect that such specialized higher level processing occurs at higher levels of the motion processing pathway. Another possible role is that the gradient itself may be an instrument for parsing the kinetic elements of a visual scene –such that slower moving components are processed at lower levels of the hierarchy while faster moving components are processed at higher levels. This is consistent with experiment. For example, Foster et al found that highly direction selective neurons in the macaque V1 and V2 were more likely to be tuned to higher temporal frequencies and lower spatial frequencies, i.e. higher speeds (Foster et al., 1985). In other words, neurons that are higher up in the V1 to MT hierarchy are more likely to be tuned to higher speeds.

Though the exact purpose of the TFFG remains to be determined, it is clear that mathematical models of visual motion neurons must encode this well-preserved emergent property. The notion of a neural basis for visual motion perception mandates that mathematical descriptions of the receptive fields of neurons located early in the pathway somehow encode the properties which emerge on a network level. In effect, the receptive fields of these early neurons are building blocks, and serve as the "DNA" of emergent properties such as TFFG. In turn, these emergent properties serve as substrates of perception precursors. This cascade places careful constraints on mathematical models, requiring them to be consistent not only with behavior observed on the single neuron level, but also with trends which are seen to emerge in population studies along the neural network. For visual motion, and in relation to the TFFG emergent property, we have shown here that the Sinc wavelet satisfies these constraints. We have argued that regarding the TFFG property, the sinc wavelet is a simpler more efficient model than existing models such as the standard spatiotemporal Gabor wavelet. Future work will include head-to-head goodness of fit – to experimental data – comparisons of the sinc function wavelet to other receptive field models.

Of note, progressive specialization along the V1 to MT motion processing stream is also observable in the morphological characteristics of the neurons. For instance, distinguishing attributes of the V1-MT or V2-MT projectors such as size, arborization patterns, terminal bouton morphology, and distribution have been observed (Anderson & Martin, 2002; Anderson, Binzegger, Martin,

& Rockland, 1998; Rockland, 1989; Rockland, 1995). Sincich and Horton demonstrated that layer 4B neurons projecting to area MT were generally larger than those projecting to layer V2 (Sincich & Horton, 2003). Overall, neurons in the V1 to MT motion processing stream are highly and progressively specialized. Figure 6 is a schematic illustration of the increased complexity in histology, waveform, and temporal frequency filtering character along the V1 to MT motion stream.

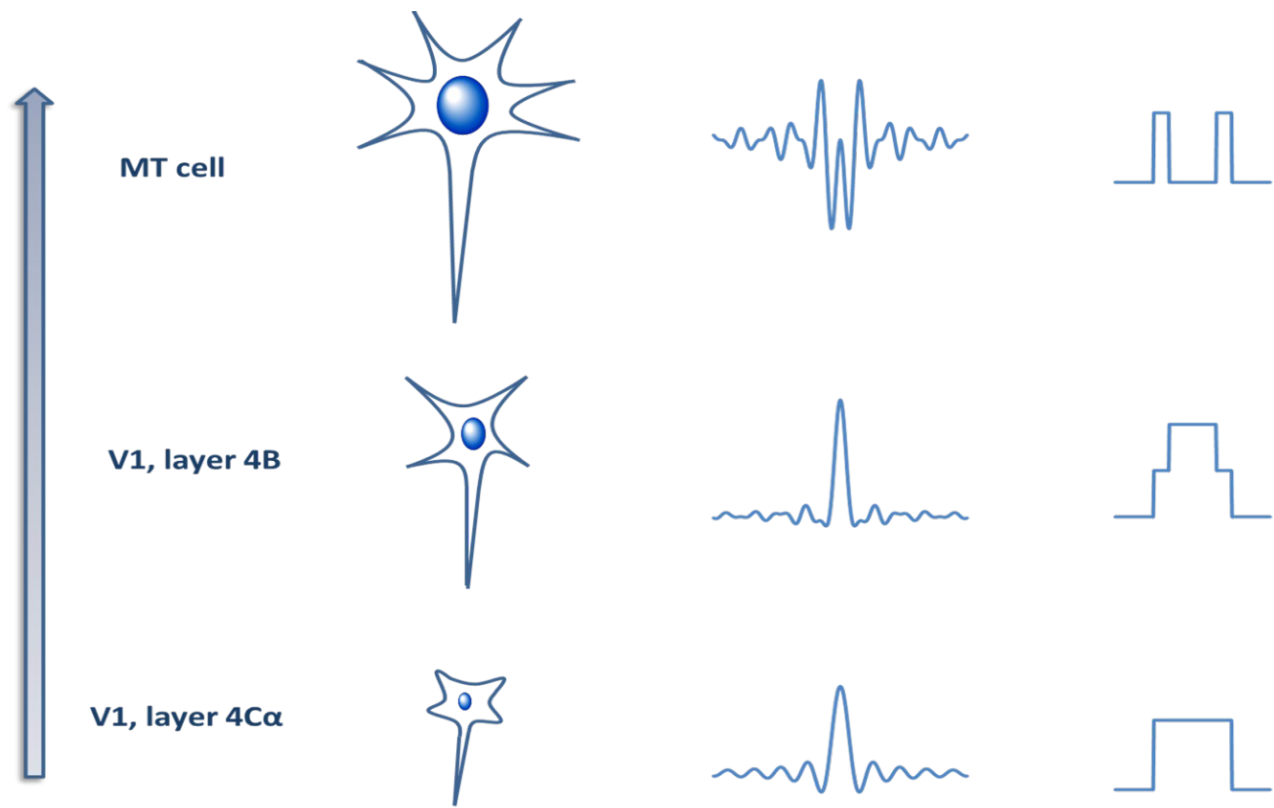

**Figure 6: Progressive complexity in histology, waveform, and temporal frequency filtering character along the V1 to MT motion processing stream.**

## CONCLUSION

Here, we have presented a new mathematical model of the receptive field of neurons in the motion cortex. This model, *the sinc wavelet model*, is the first to faithfully represent the fundamental emergent property of Temporal Frequency Filtering Gradience (TFFG). TFFG is the change in neuron temporal frequency filtering character from lowpass to bandpass along the V1 to MT network hierarchy. The sinc wavelet model will yield fundamental new insights into how the brain represents visual information.

## **ACKNOWLEDGEMENTS**

The author thanks Ari Rosenberg and Peter Blair for his helpful discussion.